\documentclass[preprint,12pt]{elsarticle}

\usepackage{amssymb}
\usepackage{amsmath}
\usepackage{float}
\usepackage{bm}
\usepackage{color}

\renewcommand*{\vec}[1]{\mathbf{#1}}

\begin{document}
\begin{frontmatter}



\title{Role of cohesion in the formation of kink wave fronts in vibrofluidized granular materials}


\author[inst1,inst2]{Han Yue}
\affiliation[inst1]{organization={Collective Dynamics Lab, Division of Natural and Applied Sciences},
            addressline={Duke Kunshan University}, 
            city={Kunshan},
            postcode={215316}, 
            country={China}}
\affiliation[inst2]{organization={University of Shanghai for Science and Technology},
            addressline={Shanghai}, 
            postcode={200093}, 
            country={China}}

\author[inst1]{Huyue Yan}
\author[inst1]{Kai Huang\corref{cor1}}
\cortext[cor1]{Corresponding author}
\ead{kh380@duke.edu}

\begin{abstract}

The formation of kink wave fronts (KWFs) in a quasi-two-dimensional granular system is investigated numerically with a focus on the role of cohesive interactions between individual particles. The cohesive particle-particle interaction is achieved through tuning the velocity-dependent coefficient of restitution, based on an analytical model introduced recently \cite{gollwitzer_2012_coefficient,mller_2016_influence}. A comparison with experimental results \cite{butzhammer_2015_pattern} indicates that the threshold for the emergence of traveling KWFs matches the regime in which the center of mass height of the granular layers fluctuates with a period that triples the vibration period. Further comparisons between dry and wet granular dynamics reveal that KWFs are more pronounced in wet granular layers because of enhanced collective motion induced by cohesion.

\end{abstract}




\end{frontmatter}


\section{Introduction}

From sand ripples to Barchan dunes \cite{Charru2013}, from density segregation and sorting \cite{Schroter2006,Groh2011} to Faraday crispation \cite{Faraday1830}, from kink to density wave fronts \cite{huang_2011_period,zippelius_2017_densitywave}, pattern formation, particularly emerging time and length scales associated with it, provides indispensable insights in deciphering the dynamics of granular materials \cite{Aguirre2021,Beiser2018,Duran2000,Jaeger1996,Brown1970}. As a matter of fact, the history of civilization is closely related to the utilization of sand, the second most abundant materials on earth \cite{Beiser2018, Duran2000}. 

As our blue planet is largely covered with water, wetting liquids always play a role in the static and dynamic behavior of granular materials in nature, industries and our daily lives. \cite{Huang2014,Tegzes2003,Herminghaus2005,Mitarai2006,Herminghaus2013}. As real-life applications typically involve a huge amount of particles, how to develop simple yet physics based model for wet particle-particle interactions is essential for large scale simulations in, e.g., process engineering or soil mechanics disciplines \cite{Li2013, Das2008}. Using phase separation as a test case, previous investigations have shown that the emergence of granular `gas bubbles' in vibrofluidized wet granular layers can be well captured with a `thin thread' minimal model considering the energy scale associated with the rupture of liquid bridges connecting adjacent particles \cite{fingerle_2008_phase} instead of the detailed force-distance relation. This is in agreement with a later investigation that shows the phase transition is indeed energy driven \cite{huang_2009_universal}. 

However, the `thin-thread' model can not reproduce pattern formation in agitated wet granular layers, such as KWFs and rotating spirals \cite{huang_2011_period}, suggesting more details in modeling liquid mediated particle-particle interactions are needed. Systematic measurements on the coefficient of restitution (CoR) for wet particle impacts show that the CoR can be analytically predicted with only two parameters: the critical Stokes number $\rm St_{\rm c}$ that characterizes a threshold inertial force for rebound to happen, as well as a characteristic CoR at infinitely large impact velocity $e_{\rm inf}$ \cite{gollwitzer_2012_coefficient, mller_2016_influence}. Although the impact model considers liquid mediated particle-particle interactions as instantaneous events, it contains the influence of inertial and viscous forces involved in the impacts, thus providing an opportunity to incorporate liquid and particle properties in molecular dynamics (MD) simulations for modeling collective behavior of wet granular materials \cite{Huang2018_model, huang_2018_internal}. 

Here, we employ the wet impact model in MD simulations to explore the collective behavior of wet granular layers under vertical vibrations. We show how the change of velocity-dependent CoR leads to the formation of KWFs through a comparison with previous experimental results.

\section{Methods}
\label{sec:setup}

Molecular dynamics simulation is employed to explore the collective dynamics of $N$ hard spheres of diameter $d = 0.8$\,mm confined in a quasi-two-dimensional container of width $L_x$ = 200$d$ and height $L_z$ = 14$d$. This configuration corresponds to a cross-section of the cylindrical container used in a previous experimental investigation\,\cite{butzhammer_2015_pattern}. The container is oscillated sinusoidally in the vertical direction with peak acceleration $\Gamma=4\pi^{2}f^{2}A/g$, where $f$, $A$, and $g$ correspond to driving frequency, vibration amplitude, and gravitational acceleration, respectively. Initially, the particles are placed randomly in the container. As simulation starts, they begin to fall freely under gravity and collide with each other as well as with the container walls. As simulation evolves, the positions of all particles are recorded every $1/(Mf)$ second with $M$=50 the number of recording frames per vibration cycle. Binary collisions are considered to be instantaneous, i.e., two particles colliding with zero CoR in the normal direction do not rebound. Nevertheless, the model only considers binary collisions, thus there are no ‘sticky’ particles in the system. The same rule applies to particle-wall collisions. This approach already simplifies liquid mediated particle-particle interactions \cite{Willett2000} into instantaneous events of binary collisions. Consequently, post-collision velocities of both particles in both normal and tangential directions can be solved analytically with momentum conservation and the definitions of normal and tangential CoR. For more details on the implementation of the wet collision model, including how to estimate the tangential CoR ($e_{\rm t}$) and how to avoid inelastic collapse, interested readers can refer to a previous numerical investigation on wave propagation in wet granular materials \cite{huang_2018_internal}.

As shown in previous investigations \cite{Brilliantov2004, mller_2016_influence}, the dependence of normal CoR on impact velocity is qualitatively different between dry and wet particle impacts. More specifically, the normal CoR for dry particle impacts is $e_{\rm dry}=1-\kappa v_{\rm n}^{\rm 1/5}$ with $\kappa =0.1$ a material property of the particles~\cite{Brilliantov2004}. While for wet particle impacts, the normal CoR is estimated with
 
\begin{equation}
\label{eq:en}
e_{\rm wet}=\left\{
\begin{array}{ll}
e_{\rm inf}(1-\rm{St}_{\rm c}/St), & \rm{St} \ge St_{\rm c} \\
0, & \mbox{otherwise}.
\end{array}
\right.
\end{equation}

\noindent where $e_{\rm inf}$ corresponds to the CoR at infinitely large impact velocity, and $\rm St_{\rm c}$ represents the critical Stokes number below which no rebound occurs. Here, the Stokes number, which measures inertial over viscous forces, is defined as $\rm St = \rho_{\rm p}dv_{\rm n}/9\eta$ with $\rho_{\rm p}$, $v_{\rm n}$ and $\eta$ particle density, particle velocity, and liquid viscosity, respectively. Based on a previous analysis~\cite{mller_2016_influence}, this dimensionless number plays a key role in determining energy dissipation arising from the flow of wetting liquid. The two parameters, namely $\rm St_{\rm c}$ and $e_{\rm inf}$, depend on the intrinsic length scales in the system (e.g., the roughness of particle surface), density ratio between particle and liquid, as well as the ratio of liquid film thickness over particle size. They are chosen (typical value $\rm St_{\rm c}=40$ and $e_{\rm inf}=0.91$) based on previous experimental measurements of the CoR for the impact of wet glass spheres~\cite{gollwitzer_2012_coefficient}. Capillary interactions are neglected here because the dynamics is dominated by the inertial and viscous forces of the liquid for the particle diameter used in the current investigation~\cite{gollwitzer_2012_coefficient}.

Note that dry CoR decreases monotonically as impact velocity increases, while wet CoR grows from 0 to $e_{\rm inf}$. As to be described below, the critical Stokes number $\rm St_{\rm c}$ plays an important role in determining the collective behavior of wet particles. The same rule is applied to particle-wall collisions. Periodic boundary condition in the $x-$ (horizontal) direction is used to avoid boundary effects. Other simulation parameters are shown in Table \ref{tab:para}. For each parameter set, a fixed simulation time of $7$\,s is chosen. Based on an analysis of granular temperature as a function of time, we found that the system reaches  its steady state within $0.5$ s (i.e., $25$ vibration cycles), therefore the simulation time, which corresponds to more than $350$ vibration cycles, is sufficiently long for the following data analysis.

\begin{table} [H]
\label{tab:para}
\centering 
\caption{\label{tab:table1} \centering Numerical values of the simulation parameters.}
\begin{tabular}{ p{5cm} p{2cm} p{3cm}  }
 \hline
 \hline
 Parameters  &  Variable  &Value\\
 \hline
 Particle density & \centering $\rho_p$   & 2500 kg $\rm m^{-3} $\\
 Particle diameter &\centering $d$ &0.0008 m\\
 Container width  &\centering $L_x$  &200$d$\\
 Container height &\centering $L_z$  &14$d$\\
 Liquid viscosity &\centering $\eta$ &1.0 mPa s\\
 \hline
 \hline
\end{tabular} 
\end{table}

\section{Kink wave fronts}
\label{sec:pd}

Under vertical vibrations against gravity, granular particles start to become mobilized upon sufficiently strong driving force, which is related to the peak acceleration applied to the system \cite{huang_2009_universal}. Depending on vibration parameters $f$ and $\Gamma$, a rich pattern formation scenario arises \cite{butzhammer_2015_pattern,zippelius_2017_densitywave} with  period-tripling rotating spiral pattern dominating \cite{huang_2011_period}. The spiral arms correspond to KWFs separating regions with different colliding phases with the container lid and bottom, giving rise to a characteristic feature to verify numerical models.

\begin{figure}[h]
	\centering
		\includegraphics[scale=.4]{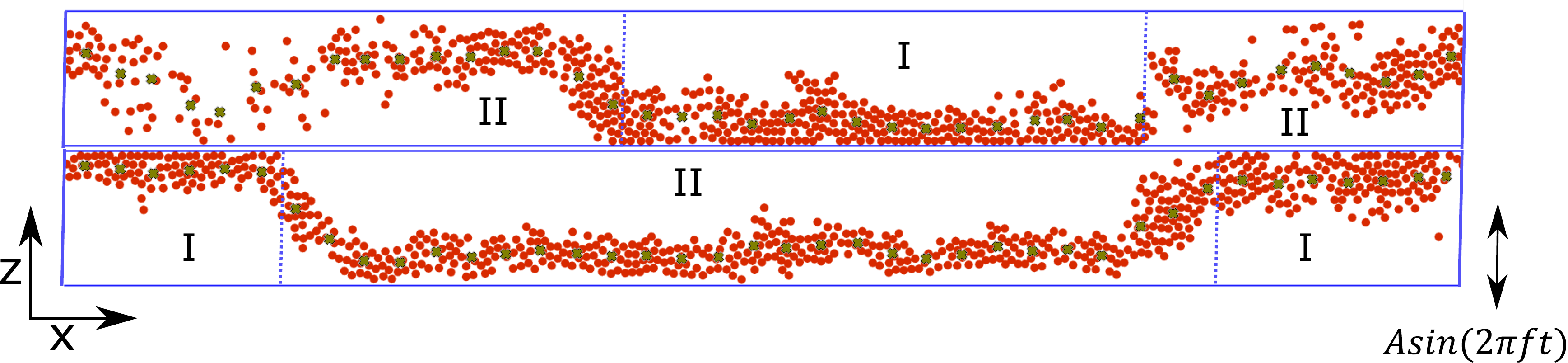} 
	\caption{Snapshots of representative simulation results for dry (upper panel) and wet (lower panel) particles in the period-tripling regime. Region I and II correspond to kink separated regions colliding with the container at different vibration cycles. Gray crosses mark the local center of mass (c.m.) height of the granular layers. Corresponding parameters are $f=70$\,Hz, $N$=700, $\rm St_{\rm c}$=40, $e_{\rm inf}$=0.91, $\Gamma$=30.}
	\label{fig:kink}
\end{figure}

Figure \ref{fig:kink} shows a comparison between dry and wet particles in the period-tripling regime, meaning that the center of mass (c.m.) of the granular layers fluctuates with a period of $3/f$. It clearly suggests that, for a system with large aspect ratio ($L_x/L_z$), the oscillation of a granular layer is unlikely to be coherent across the container. Here, $x-$ and $z-$ directions correspond to the horizontal and vertical directions perpendicular and along gravity, respectively. As the periodicity is greater than one, a separation of colliding phases with the container emerges, leading to the formation of kink waves. For dry particles, the granular layers are dilute, particularly in the free-flying region II, shortly after collisions with the container bottom. For wet particles under the same simulation conditions, the granular layers are condensed, giving rise to a clear kink front due to cohesion introduced via wet CoR. To facilitate further analysis on local fluctuations, the container is divided horizontally into $N_{\rm c} = 50$ compartments. The gray crosses mark the c.m. height of all particles in the corresponding compartment. The distribution of local c.m. suggests that wetting leads to more coherent motion.

\begin{figure}[ht]
	\centering
		\includegraphics[scale=.6]{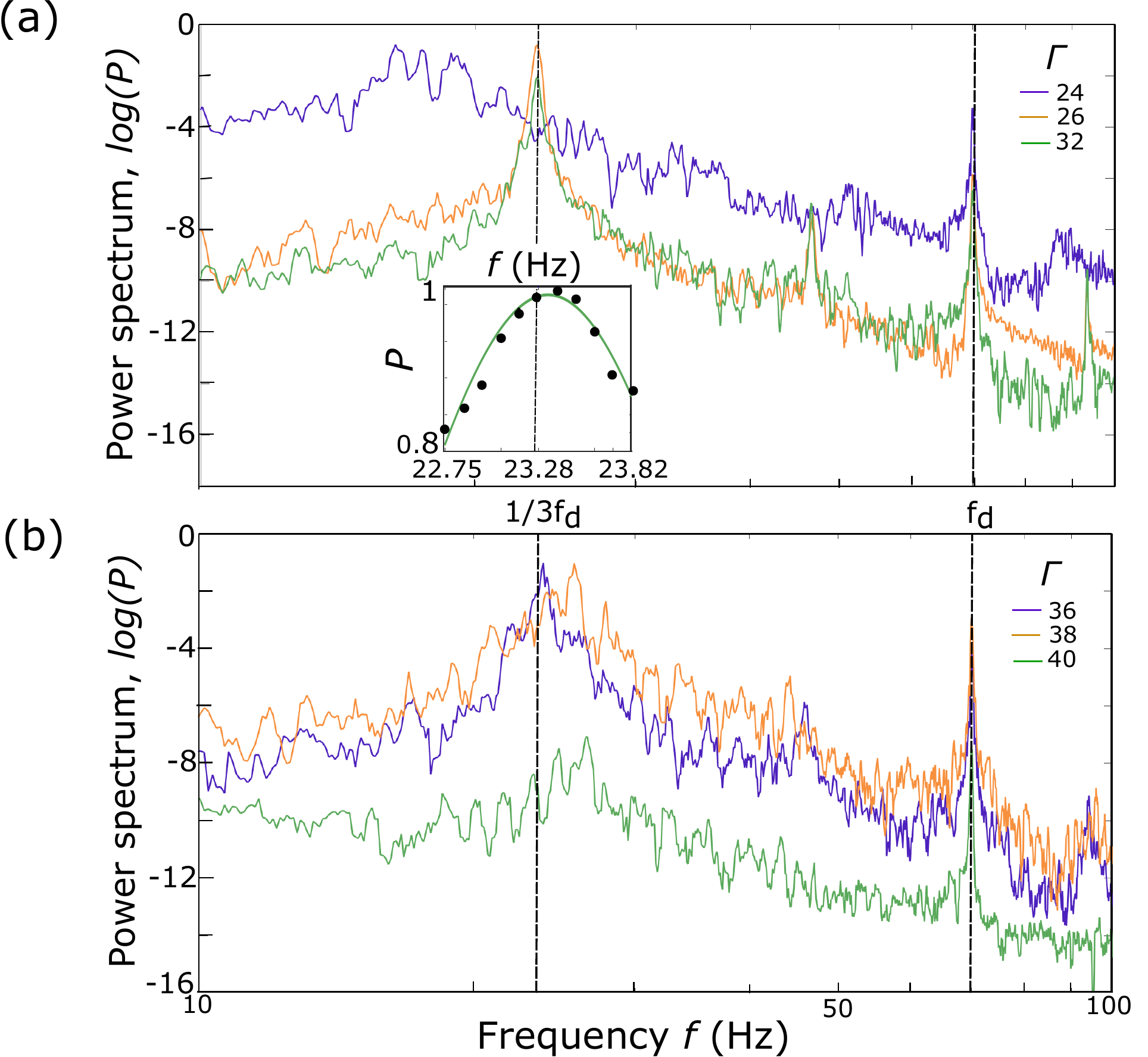} 
	\caption{Normalized power spectrum $P$ of c.m. height in a wet granular layer at lower (a) and upper (b) boundary of the period-tripling regime, as indicated by the peak at $f_{\rm d}/3$. Different colors correspond to different $\Gamma$. Other parameters are the same as in Fig.\,\ref{fig:kink}. The inset in (a) shows a close view of the fitting curve determining the peak of $P$ at $f_{\rm d}/3$.}
	\label{fig:ps}
\end{figure}

As the first step, we analyze height fluctuations with time for the whole granular layer to determine the periodicity of the system. As shown in Fig.\,\ref{fig:ps}, there exists a clear peak at driving frequency $f_{\rm d}=70$\,Hz, arising from periodic collisions with the container. To better present experimental conditions, the parameters for wet CoR (i.e,, $\rm St_{\rm c}$ and $e_{\rm inf}$) are chosen to be the same as the experimentally measured values for glass beads \cite{gollwitzer_2012_coefficient}. In addition to the pronounced peak at the driving frequency, a second peak emerges at $f_{\rm d}/3$ along with the formation of KWFs shown in Fig.\,\ref{fig:kink}(b).

Assuming the whole granular layer as a single particle colliding completely in-elastically with the container, previous work indicates that the granular layer will touch the lid of the container at $\Gamma\approx 20$ \cite{butzhammer_2015_pattern}. As $\Gamma$ increases further, additional energy injection through collisions with the container lid assists in stabilizing KWFs. This feature is clearly indicated in Fig.\,\ref{fig:ps}, where a comparison of power spectrum for various $\Gamma$ is presented to illustrate the transitions into (a) and out of (b) the period-tripling regime where KWFs dominate.    

As shown in Fig,\,\ref{fig:ps} (b), the peak at $f_{\rm d}/3$ drifts slightly to higher frequencies while becoming flattened systematically as $\Gamma$ increases. Above the upper bound of period-tripling regime, KWFs are gradually replaced by coherent motion, presumably owing to the additional confining force from the container. We note that the emerging period-tripling regime exists for the c.m. fluctuations of dry particles as well, suggesting that the single particle model \cite{butzhammer_2015_pattern} acts as a good first-order approximation for the collective dynamics.

In order to determine the threshold for period-tripling regime, we fit the peak at both $f_{\rm d}$ and $f_{\rm d}/3$ with normal distributions. Taking advantage of the prominent peak at $f_{\rm d}$ as a reference, we use the following criteria to identify period-tripling regimes: The standard deviation of the fit at $f_{\rm d}/3$ is less than twice that of the fit at $f_{\rm d}$ and the center frequency from the fit should be $\leq f_{\rm d}/3\pm 1$\,Hz. Here, the threshold is chosen to be the relative standard deviation because the peaks at $f_d/3$ is typically wider than that at $f_d$, and the boundary for period-tripling regime measured with this threshold matches well visual inspections. Using the criteria described above, it is straightforward to identify regimes with other periodicity as well.

\begin{figure}[ht]
	\centering
		\includegraphics[scale=.4]{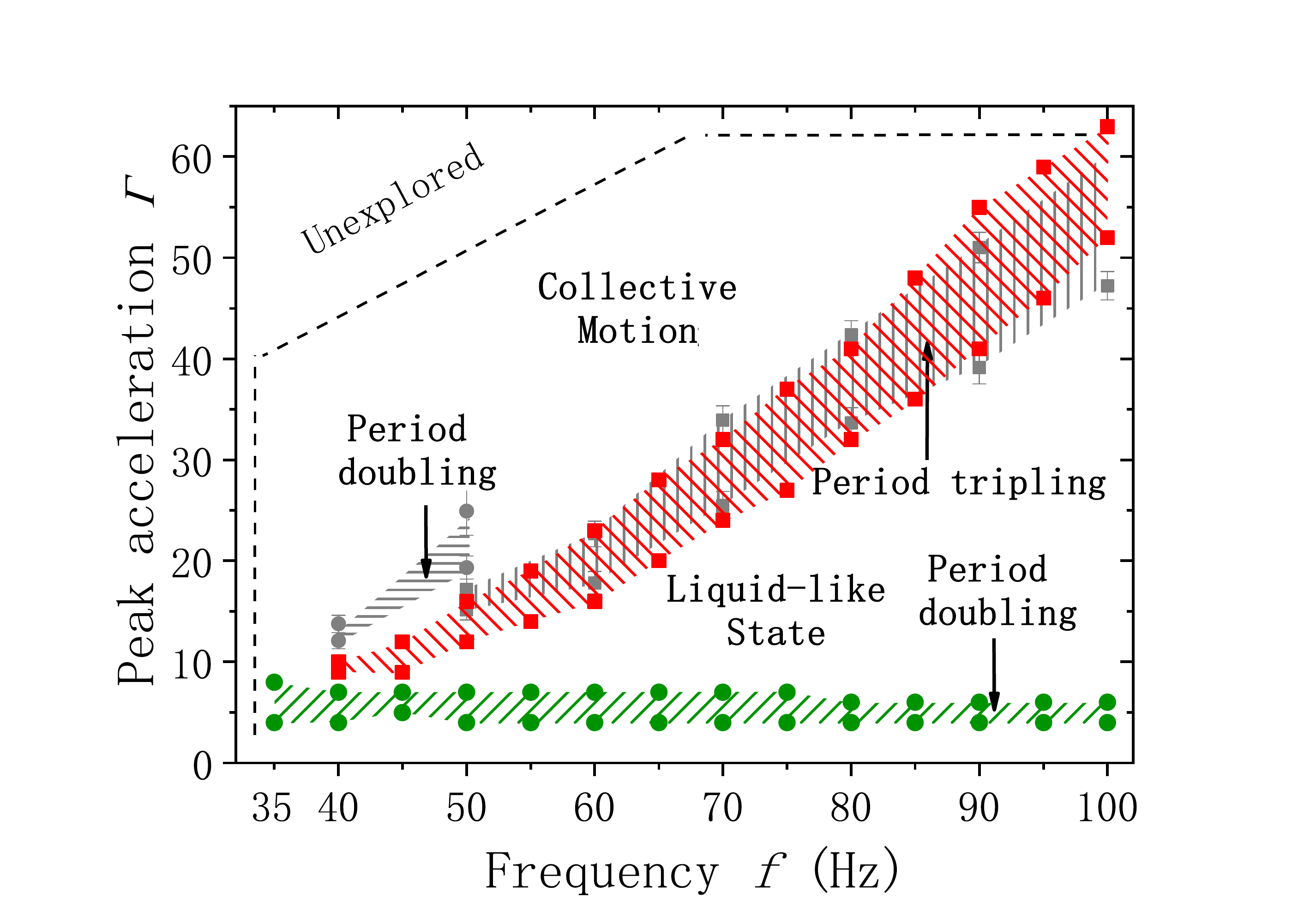}
	\caption{Stability diagram showing the collective behavior of wet particles obtained from simulations, in comparison with previous experimental data. Green and red shaded regions correspond to the period-doubling and -tripling in simulations. Green and red symbols correspond to the thresholds obtained with spectrum (see text for more details). Gray shaded regions correspond to the previous experimental results adapted from\cite{butzhammer_2015_pattern}.}
	\label{fig:pd}
\end{figure}

Figure~\ref{fig:pd} shows the stability diagram of wet granular layers under vertical vibrations, obtained using the protocol describe above. The period-doubling regime appearing at small $\Gamma$ depends weakly on driving frequency, in agreement with the previous analytical prediction $9.1$ as well as the single particle bouncing model $8.2-10.2$\,\cite{butzhammer_2015_pattern}. It indicates that the period-doubling regime arises from the free flying period of the granular layers between subsequent collisions with container bottom, before the highest granular layer reaches the lid. Experimentally, the period-doubling regime cannot be identified from top-view images because it does not yield any KWF.   

In alignment with the single particle model, a frequency dependent period-tripling regime is observed at larger $\Gamma$, as indicated by the red-shaded region in Fig.\,\ref{fig:pd}. According to the numerical results, this regime starts at $f\approx 40$\,Hz and expands as $f$ increases. It agrees well with previous experimental results with the same free space above granular layers (i.e., container height subtracted by the granular layers thickness) for $f \ge 60$\,Hz, because the free space determines the timescale for granular layers to fly freely \cite{butzhammer_2015_pattern}. Different from the experimental results, numerical simulations reveal an extended period-tripling regime to lower frequency $f \approx 40$\,Hz and $\Gamma\approx 10$. This feature arises presumably from the period-tripling regime emerging before the highest granular layer collides with the container lid. As $\Gamma$ increases further at low $f$, a period-doubling regime (horizontally shaded in gray) emerges in experiments because of the presence of KWFs. However, this period-doubling regime is absent in simulations, suggesting additional conditions for the standing KWFs. Suppose there are two segments of granular layers (i.e., two KWFs) colliding with the container in subsequent vibration cycles. The horizontal momentum generated via one segment colliding with the container should be comparable to that from the other segment. In simulations, the probability for the granular layers to split into two equal segments is low, which could be one possible reason for the lack of period tripling KWFs in simulations. Further investigations in this direction shall help optimize the impact model.

In short, the above comparisons suggest that period-tripling regime identified by the height fluctuations of wet granular layers matches the experimental results quantitatively. On the one hand, it demonstrates that period-tripling is an essential condition for the emergence of traveling KWFs. On the other hand, it also triggers the question on the role of cohesion played in the formation of KWFs. Note that traveling KWFs are only observed in wet granular layers, although both dry and wet granular layers yield the same threshold for the period tripling regime \cite{butzhammer_2015_pattern}.

\section{Influence of cohesion}

\label{sec:wetting}

\begin{figure}[H]
\centering
 \includegraphics[scale=.55]{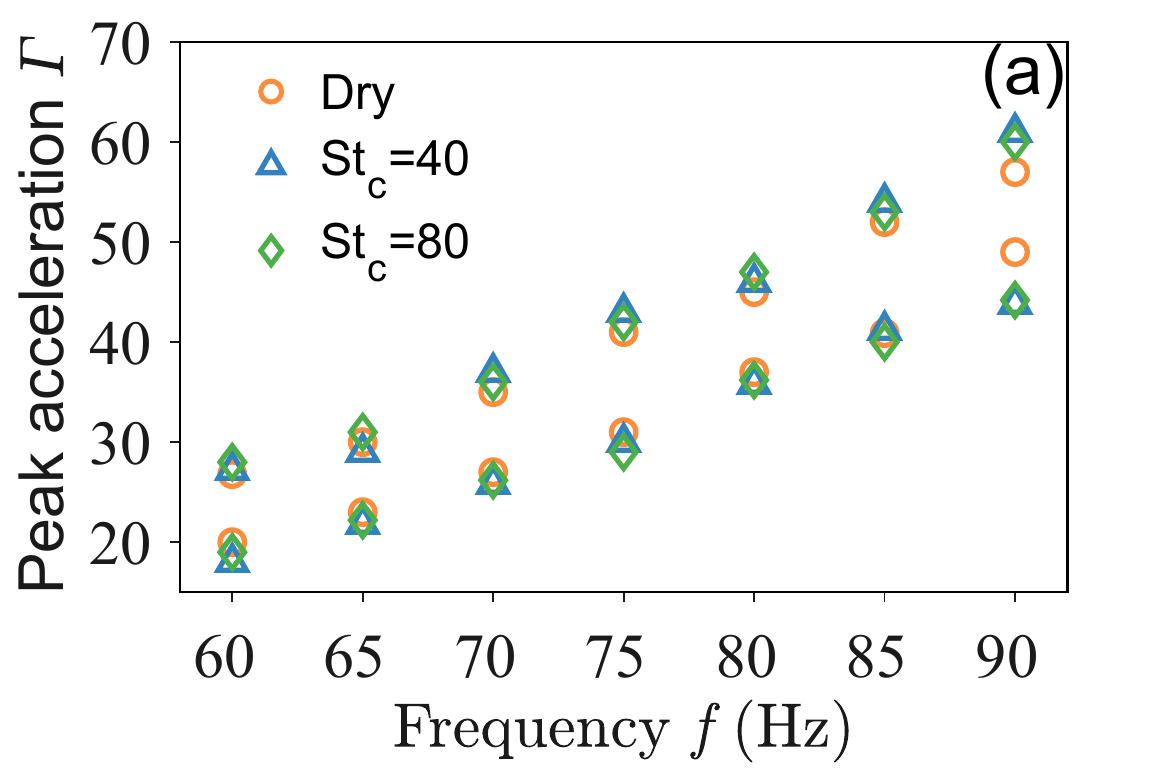} 
\quad
 \includegraphics[scale=.55]{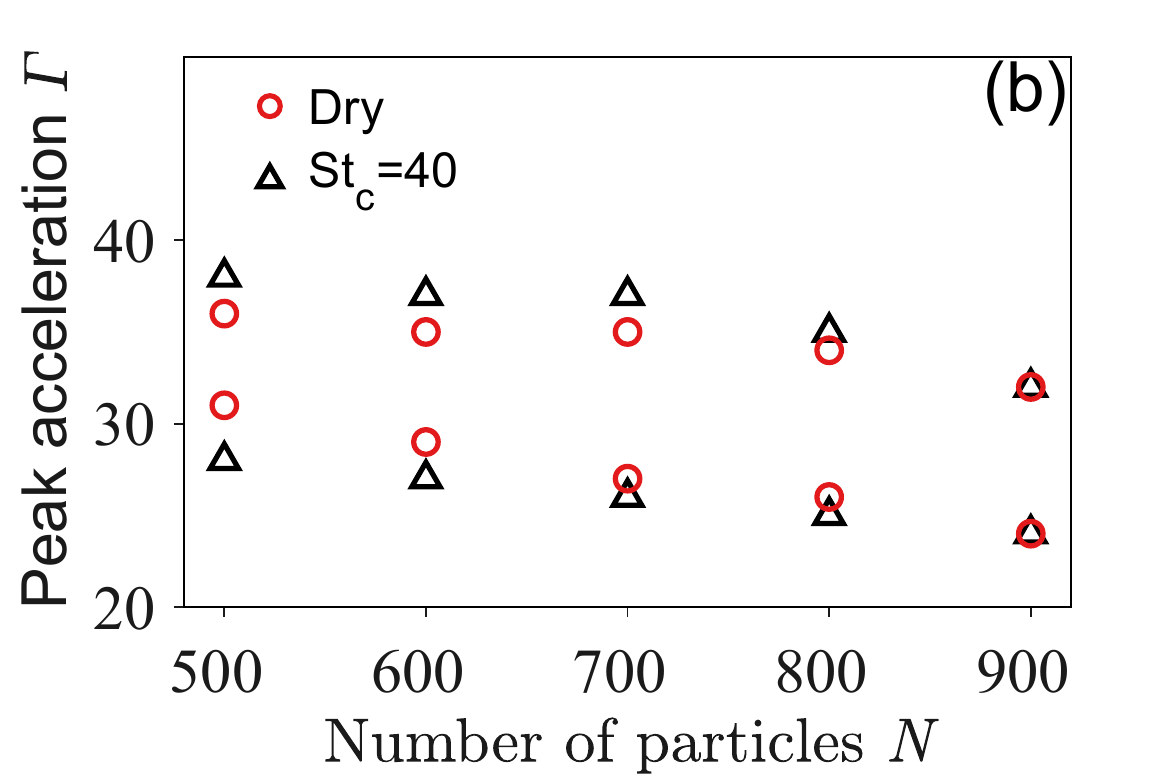} 
 \includegraphics[scale=.55]{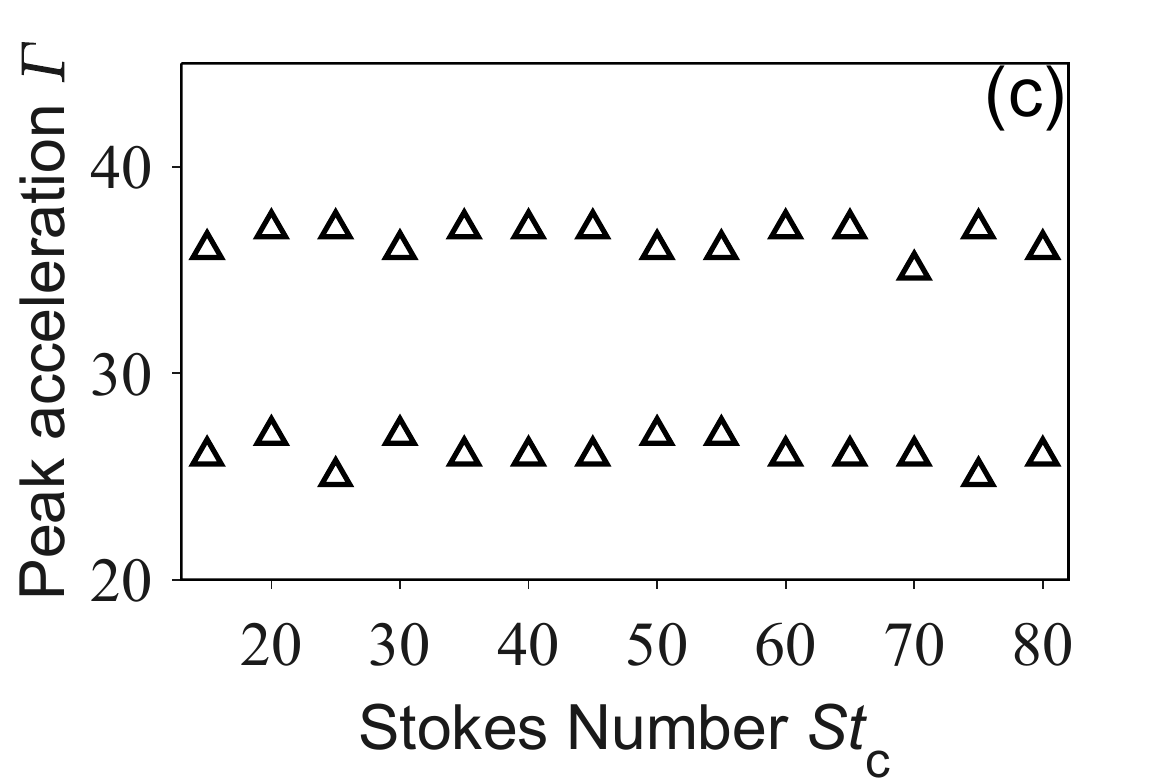} 
\quad
 \includegraphics[scale=.55]{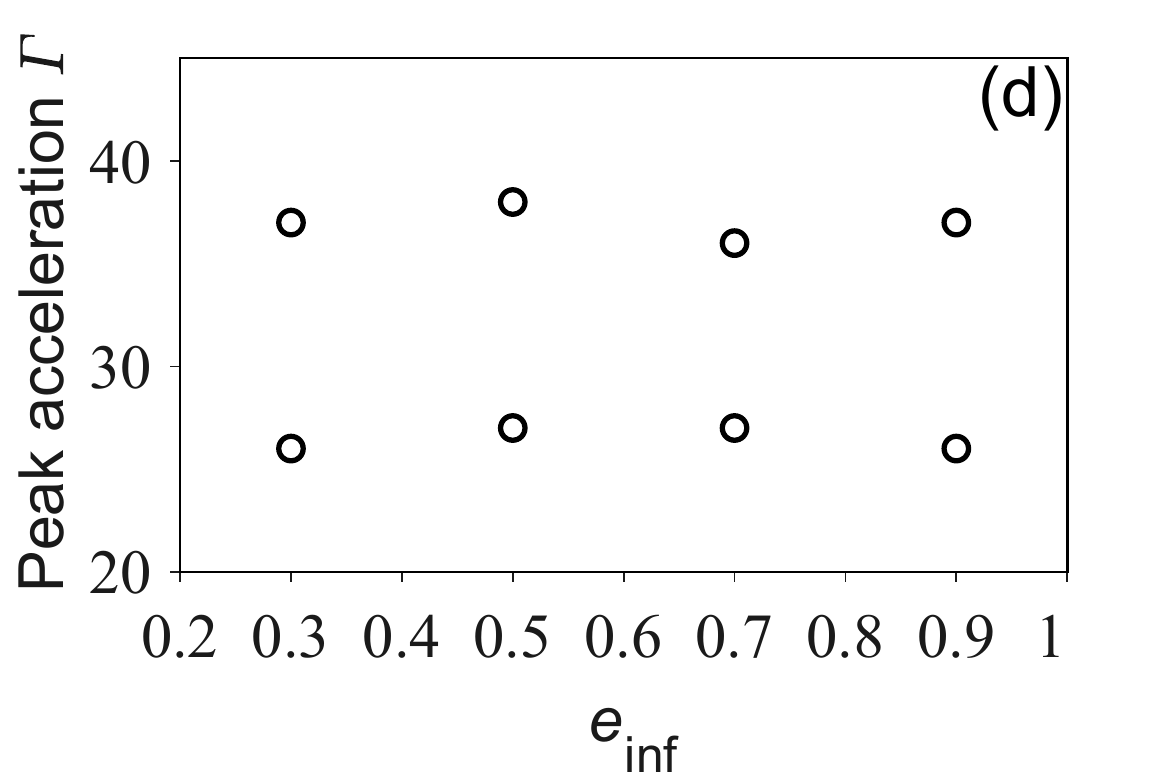} 

\caption{Dependence of period-tripling regime on various simulation parameters: (a) Comparing dry and wet particles with fixed $N=700$ and $e_{\rm inf}=0.91$; (b) Varying particle number $N$ at fixed $f=70$\,Hz, $St_{\rm c}=40$, and $e_{\rm inf}=0.91$; (c) Varying $St_{\rm c}$ at fixed $f =70$\,Hz, $N=700$, and $e_{\rm inf}=0.91$;(d) Varying $e_{\rm inf}$ at fixed $f =70$\,Hz, $N=700$, and $St_{\rm c}=40$.}
\label{fig:para}
\end{figure}

The above analysis suggests that period-tripling driven KWFs represent the collective motion of the whole granular sample as a point mass, thus the boundary should be weakly dependent on the wetting condition. For the extreme case of completely inelastic collisions for all particles (i.e., extremely large $\rm St_{\rm c}$), the mean granular temperature further decreases and consequently particles move coherently as a single particle colliding completely inelastically with the container. In order to verify this statement, we conduct a parametric study with systematically tuned simulation parameters associated with wetting. 

As shown in Fig\,\ref{fig:para}(a), the boundaries of period-tripling regime for wet and dry particles overlap with each other very well except for the difference at $f=90$, indicating that particle-particle interactions do not play an important role in determining the period-tripling regime. At $f=90$\,Hz, the coherent motion for dry granular layers is easier to be disturbed due to frequent collisions with the container. 

Figure~\ref{fig:para}(b) shows the dependence of the thresholds on the number of particles $N$ in the granular layers. Generally speaking, granular particles, no matter dry or wet, tend to cluster (i.e., move collectively) because of dissipative particle-particle interactions. Therefore, the growth of $N$ results in higher energy dissipation and consequently lower threshold for the period tripling regime to emerge, in agreement with experiments \cite{butzhammer_2015_pattern}. This is reminiscent to the influence of $f$ discussed above, both decreasing particle number (i.e., lower collision rate) and increasing the driving frequency (i.e., more frequent disturbance from driving) lead to enhanced random motion. For particles with a higher tendency to move randomly, entering period-tripling regime requires lower $\Gamma$ for wet than for dry granular layers because of the enhanced cohesion.

The influence of control parameters ($\rm St_{\rm c}$ and $e_{\rm inf}$) for wet CoR is presented in Fig.\ref{fig:para}(c) and (d). A variation of $\rm St_{\rm c}$ from $15$ to $90$  yields the same threshold within the experimental uncertainty. Similarly, the threshold $\Gamma_{\rm c}$ is found to be weakly dependent on $e_{\rm inf}$.

Based on the above analysis, we conclude that wetting will not influence the threshold for the period-tripling regime, which arises from the collective motion of the whole granular layers as a single particle. Consequently, it is intuitive to ask why rotating spirals or propagating KWFs are not observed in dry granular layers under the same driving conditions. In other words, which role does wetting liquid play in pattern formation?

\begin{figure}[H]
   \centering
     \begin{minipage}[b]{0.45\linewidth}
     \includegraphics[scale=.27]{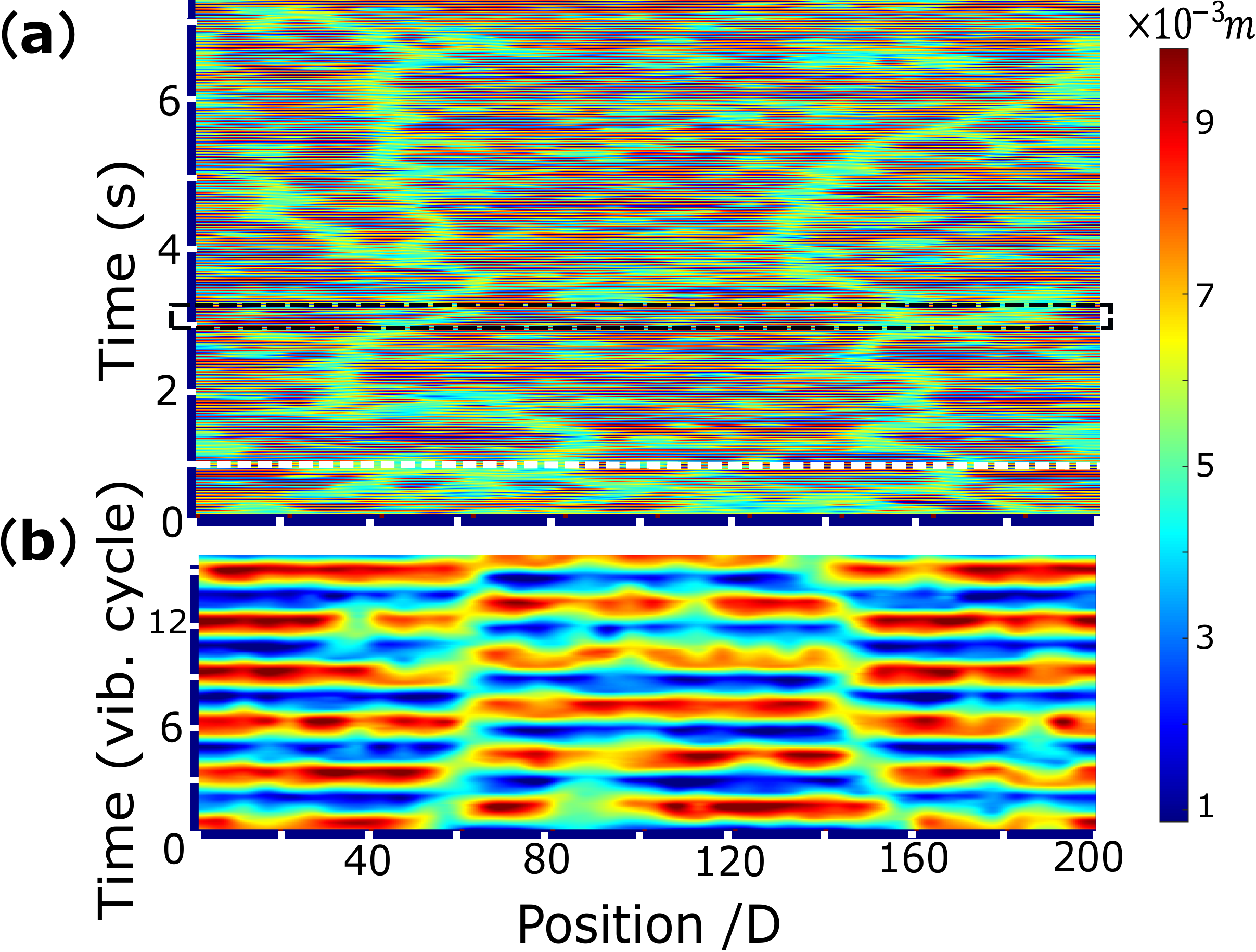}
     \label{fig:minipage1}

   \end{minipage} 
   \quad
  \begin{minipage}[b]{0.45\linewidth}
      \includegraphics[scale=.27]{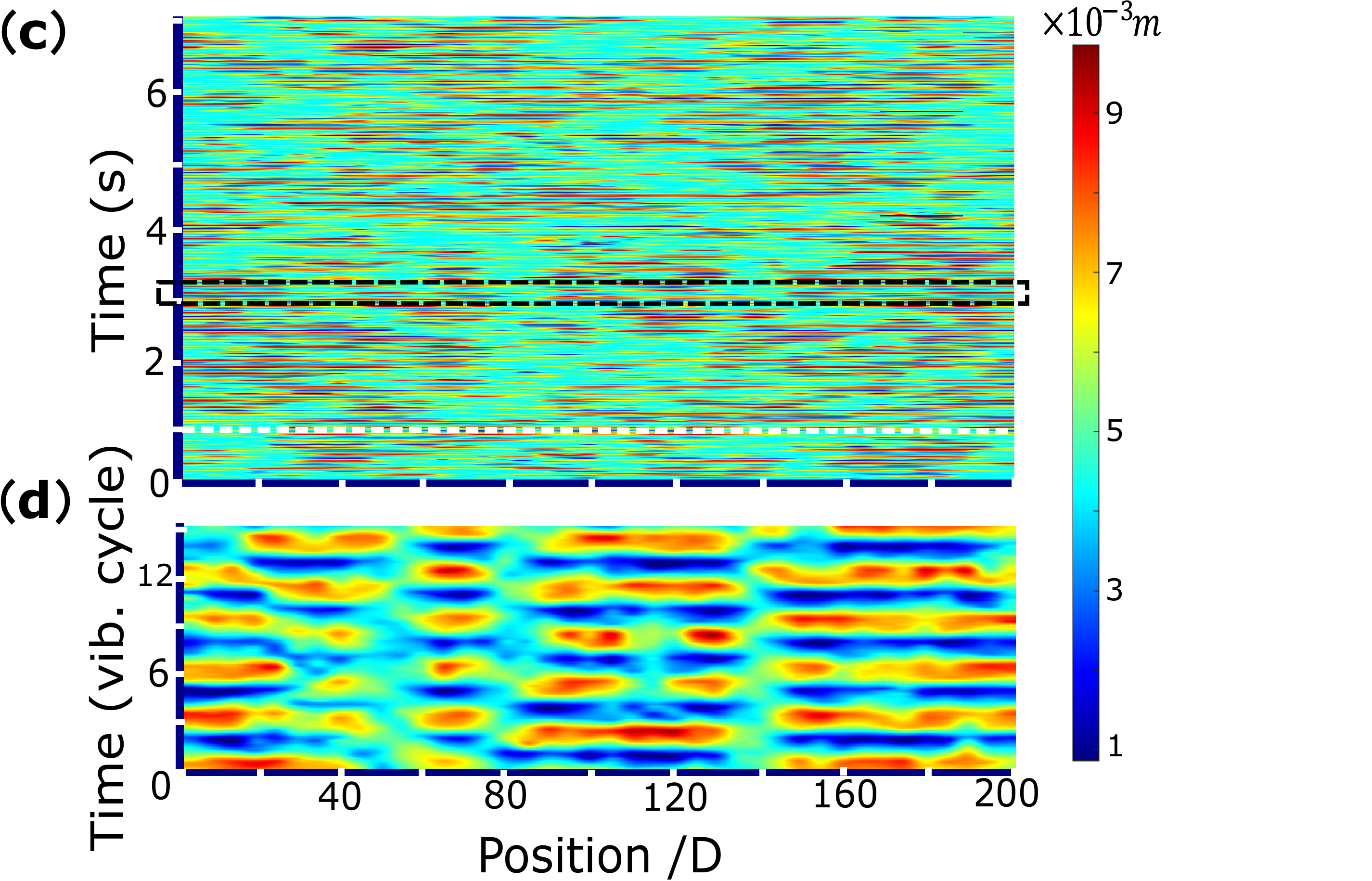}
    \label{fig:minipage2}

    \end{minipage} 
    \caption{Time-space plots of c.m. height in various horizontal compartments as a function of time for both wet (a, b) and dry (c, d) particles. (a) and (c) correspond to the long-time (up to a few seconds) behavior of height fluctuations, while (b) and (d) correspond to zoom-in views of the fluctuations within a few vibration cycles (black dashed boxes marked in the upper panels). Simulation parameters are the same as in Fig.\,\ref{fig:kink}}
    \label{fig:hmap}
\end{figure}

To investigate the difference between wet and dry particles, the time-space plots of c.m. height fluctuations are presented in Fig.\,\ref{fig:hmap}. For both dry and wet cases, period tripling behavior is clearly visible in the zoom-in view of c.m. fluctuations shown in (b) and (d). A KWF corresponds to locations where a 'kink' of height (i.e., where c.m. height persists) exists. As time evolves, the location of the `kink' propagates with time. However, the emergence and propagation of KWFs are clearly visible in wet granular layers, not in dry ones. More specifically, two KWFs emerge at about $1$\,s (white dashed line) after the system is initialized. Subsequently, the KWFs meander back and forth. One front bifurcates at about $3.5$\,s. One of the branches disappears shortly at about $4.5$\,s, while the other one persists. Zoom-in view in the lower panels  obvious indicates that the KWF separates regions with different heights. The clear contrast between c.m. height around the kink front region in Fig.\,\ref{fig:hmap}(b) in comparison to (d) illustrates the influence of cohesion on the formation of KWFs. For dry particles under the same driving conditions, no persistent kink waves are identified [see Fig.\,\ref{fig:hmap}(c), same color code is used for all subplots]. Although c.m. fluctuations still have period-tripling behavior [see Fig.\,\ref{fig:hmap}(d)], the weaker contrast indicates less coherent motion for dry than for wet granular media.

\begin{figure}[ht]
	\centering
		\includegraphics[scale=0.7]{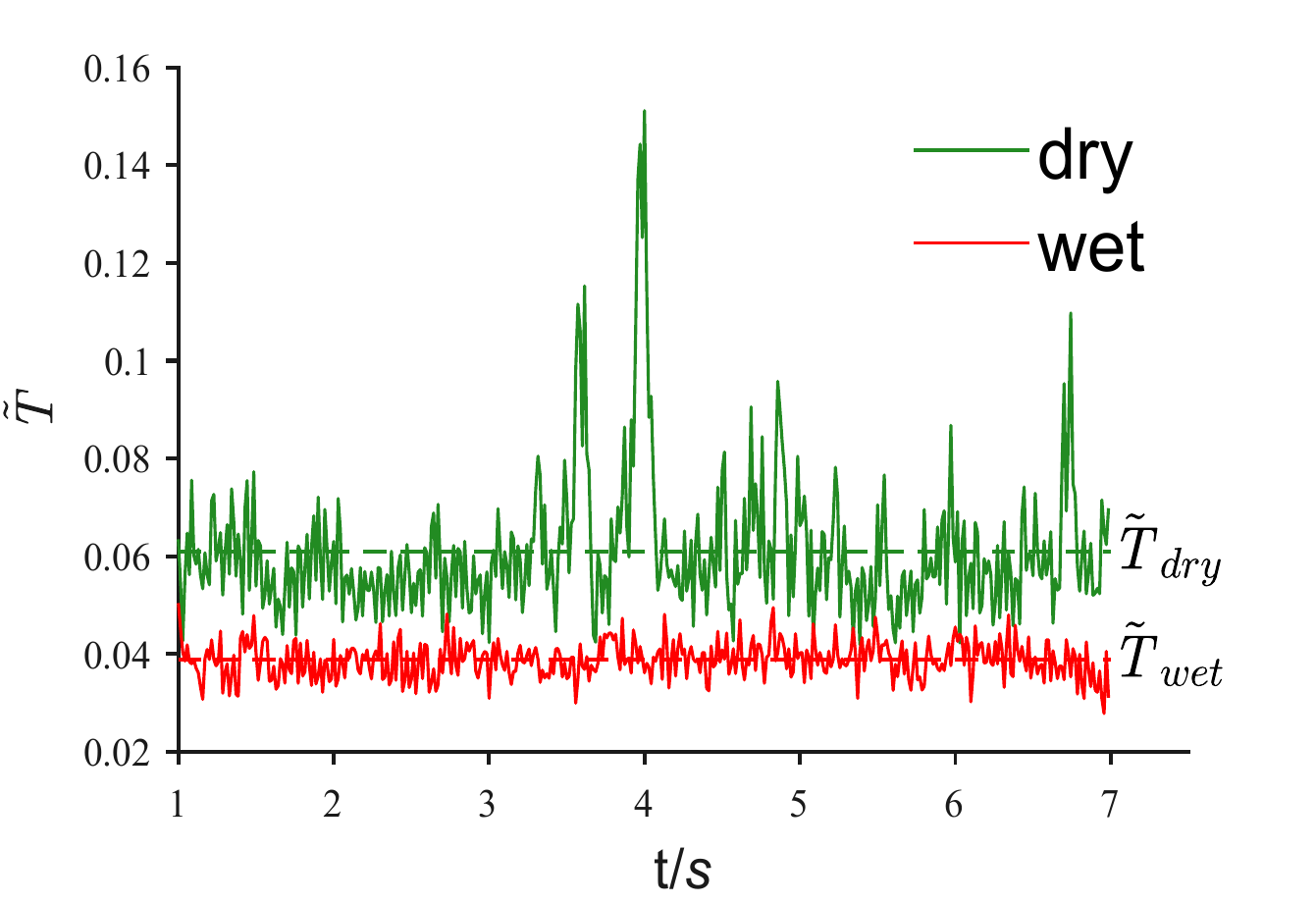}
	\caption{Rescaled granular temperature $\widetilde{T}$ as a function of time for both wet and dry samples in the steady state. Dashed lines are the mean values of dry $\widetilde{T}_{\rm dry}$ and wet $\widetilde{T}_{\rm wet}$ granular temperature as a function of time. Other parameters are the same as in Fig.\,\ref{fig:kink}}
	\label{fig:T}
\end{figure}

In order to quantify the coherent motion of granular layers in the period-tripling regime, we use rescaled granular temperature $\widetilde{T}$ as a measure for the `randomness' of particles in both dry and wet cases. Rescaled granular temperature is defined as $\widetilde{T}=\bar{T}/v_0^2$, where $v_0=\Gamma g/{(2\pi f)}$ is the peak velocity of the vibrating container. The average granular temperature is defined as $\bar{T} = \sum_{j=1}^{N_c} T_j / N_c$  with $T_j$ granular temperature measured in the $j$th compartment. Note that it is necessary to sample granular temperature in each individual compartment because collective motion of granular layers in different compartments is not always in phase (see the lower panels of Fig.\,\ref{fig:hmap}). Here, `randomness' of particles in the compartment $j$ is characterized with granular temperature $T_j = \sum_{i=1}^{N_i}(\vec{v}_{ij}-\bar{\vec{v}}_j)^2 / N_i$ with $N_i$ the number of particles in the $j$th compartment. The mean velocity of all particles in compartment $j$ is $\bar{\vec{v}}_j=\sum_{i=1}^{N_i}\vec{v}_{ij} / N_i$, where $\vec{v}_{ij}$ corresponds to the velocity vector of $i$th particle in compartment $j$. As shown in Fig.\,\ref{fig:T}, granular temperature in dry granular layers ($0.061$) is about $50$\% larger than that in wet granular layers ($0.039$). $\widetilde{T}_{\rm dry}$ exhibits much stronger fluctuations than $\widetilde{T}_{\rm wet}$. Both features clearly demonstrate enhanced collective motion induced by wetting. This comparison suggests that the enhanced collective motion plays an essential role in the formation of KWFs in wet granular layers. For the case of dry granular layers, although period tripling regime exists in the stability diagram shown in Fig.\,\ref{fig:pd}, stronger fluctuations of granular temperature due to collisions with the vibrating container effectively prohibit the emergence of persistent KWFs. Therefore, we conclude that cohesion induced coherent motion of granular layers in a period-tripling fashion leads to the formation of traveling KWFs.

\section{Conclusion}

To summarize, we show that a simplified model previously proposed for wet particle collisions\,\cite{mller_2016_influence} can be used to reproduce propagating KWFs in vibrofluidized wet granular layers\,\cite{butzhammer_2015_pattern}. The quantitative agreement of the period-tripling regime in the stability diagram obtained numerically with that from experiments indicates that KWFs emerge as a consequence of period-tripling bifurcation of the collective motion of granular layers as a whole. Further parametric studies suggest that periodicity arises from the collective motion of the granular layers, regardless of the type of interactions between neighboring particles. Further parametric studies suggest that the qualitative difference between dry and wet particle-particle interactions determines whether KWFs emerge or not. In other words, the detailed nature of wet particle impact, such as the amount of energy loss in each collision, does not play a dominating role in the formation of KWFs. This investigation suggests the possibility of implementing the simplified wet impact model for understanding the collective dynamics of wet granular materials. In the future, further verification of the wet impact model through comparisons to other instabilities observed in the model system are needed before applying it more broadly.

\section{Acknowledgement}
We thank Jinchen Zhao, Chen Lyu, and Simeon V\"olkel for helpful discussions. This work is partly supported by the Deutsche Forschungsgemeinschaft through Grant No.~HU1939/4-1 and Startup Grant from Duke Kunshan University.


\bibliographystyle{elsarticle-num} 





\end{document}